\documentclass[12pt,a4paper]{article}
\input epsf
\usepackage{amsmath}
\usepackage{amssymb}

\newcommand{\pint}{\makebox[0pt][l]{\hspace{3.1pt}$-$}\int}
\newcommand{\ellK}{{\rm K}}

\newcommand{\ellE}{{\rm E}}

\begin{document}
\thispagestyle{empty}
\begin{flushright}
{\sc\footnotesize hep-th/0401057} \\
{\sc\footnotesize AEI-2004-001}\\
{\sc\footnotesize SPhT-T04/002}
\end{flushright}
\vspace{1cm}
\setcounter{footnote}{0}
\begin{center}
{\Large{\bf Planar \mathversion{bold}${\cal N}=4$ Gauge Theory 
and the Inozemtsev Long Range Spin Chain \par}}\vspace{20mm}
{\sc Didina Serban$^a$ and Matthias Staudacher$^b$} \\[7mm]
$^a${\it Service de Physique Th\'eorique, CNRS-URA 2306
\\
C.E.A.-Saclay \\
F-91191 Gif-sur-Yvette, France}\\[1mm]
$^b${\it Max-Planck-Institut f\"ur Gravitationsphysik\\
     Albert-Einstein-Institut \\
     Am M\"uhlenberg 1, D-14476 Potsdam, Germany} \\ [2mm]
{\tt serban@spht.saclay.cea.fr, matthias@aei.mpg.de} \\[10mm]
{\sc Abstract}\\[2mm]
\end{center}

\noindent We investigate whether the (planar, two complex scalar) 
dilatation operator of ${\cal N}=4$ gauge theory can be,
perturbatively and, perhaps, 
non-perturbatively, described by an integrable long range spin chain
with elliptic exchange interaction. Such a chain was introduced
some time ago by Inozemtsev. In the limit of sufficiently ``long'' 
operators a Bethe ansatz exists, which we apply at the perturbative
two- and three-loop level. Spectacular agreement is found with spinning
string predictions of Frolov and Tseytlin for the two-loop energies 
of certain large charge operators. However, we then go on to show
that the agreement between perturbative gauge theory and 
semi-classical string theory
begins to break down, in a subtle fashion, at the three-loop level.
This corroborates a recently found disagreement between
three-loop gauge theory and near plane-wave string theory results,
and quantitatively explains a previously obtained puzzling deviation between the string proposal and a numerical extrapolation of finite
size three-loop anomalous dimensions.  
At four loops and beyond, we find that the Inozemtsev chain exhibits
a generic breakdown of perturbative BMN scaling. However,
our proposal is not necessarily limited to perturbation theory,
and one would hope that the string theory results can be recovered
from the Inozemtsev chain at strong 't Hooft coupling.

\newpage

\setcounter{page}{1}

\section{Introduction and conclusions}

Nearly two years ago a fresh approach \cite{bmn} to uncover the dynamics
of the AdS/CFT correspondence \cite{ads} was proposed. 
Progress in this direction is clearly required if we are to understand
the gauge theory implications of the correspondence for quantum gravity
and string theory, as well as the string theory implications of AdS/CFT
for gauge theory. Berenstein, Maldacena and Nastase (BMN) \cite{bmn}
argued that the string spectrum on $AdS_5 \times S^5$ in a 
linearized (plane wave) limit, which is known \cite{metsaev},
when expressed in planar ${\cal N}=4$ gauge theory language,
leads to a prediction for the all-loop exact anomalous dimensions
of certain operators containing one large R-symmetry charge $J$.
These predictions take the form of explicit functions 
in the BMN coupling constant 
\begin{equation}
\label{eq:lambda'}
\lambda'=\frac{\lambda}{J^2},
\end{equation}
which are analytic around $\lambda'=0$, with a finite radius of
convergence. One can therefore expand them in integer power series
in $\lambda'$, and thus, so it seems, $\lambda$. This led BMN to suggest
that the predictions could be checked, order by order,
in {\it perturbative} planar ${\cal N}=4$ gauge theory.
The claim is that thereby one avoids a notorious AdS/CFT
difficulty: the regime where AdS string theory is calculable
corresponds to $\lambda$ large, while gauge theoretic
perturbation theory obviously assumes $\lambda$ to be small.
We will have more to say about this below.

The BMN proposal triggered a large number of interesting
research papers, most of which cannot be discussed here (see the
thorough reviews \cite{bmnreview}). In particular, powerful
techniques for the efficient computation of one-loop anomalous
dimensions of ${\cal N}=4$ composite operators containing
a large number of scalar fields were developed
\cite{effective}. These activities led to the highly important
insight of Minahan and Zarembo \cite{mz} that the {\it planar} one-loop
anomalous dimensions of these operators can be found
by diagonalizing the Hamiltonian of an equivalent
integrable spin chain by the Bethe ansatz method. 
In the case of just two complex scalars the spin chain is
extremely simple, the Heisenberg XXX model. This is the
``harmonic oscillator'' of condensed matter theory. Similar
spin chain techniques had been previously developed in the QCD context
for a different class of operators, rather related to the
space-time symmetries instead of the R-symmetry of the
scalar operators \cite{lipatov}. 
A unified one-loop treatment of {\it all} conformal
operators in the ${\cal N}=4$ theory was developed in 
\cite{b1}, and shown to yield an integrable 
PSU$(2,2|4)$ super spin chain \cite{bs}, as well as its corresponding
Bethe ansatz.

The discovery of integrability in perturbative
${\cal N}=4$ theory is, potentially, of great importance. In a parallel
development integrable structures were also pointed out on
the string side of the correspondence \cite{wadia}. Since the
structures in gauge and string theory appear in very
different regimes, the exciting prospect arises that
the planar AdS/CFT system might be completely and non-perturbatively
integrable. If true, this might allow to ``exactly solve'' planar
${\cal N}=4$ gauge theory, and thus free IIB string theory on
$AdS_5 \times S^5$.

One way of looking at the BMN approach is to consider it a 
semi-classical limit \cite{gkp} with an ``artificial'' large parameter, the
charge $J$. In fact, a beautiful generalization of the BMN limit
was proposed by Frolov and Tseytlin \cite{ft2} (see also \cite{ft0},
and the closely related earlier work \cite{earlier}). 
They consider rotating string solutions
of the classical string equations of motion with {\it several} large 
charges $J_1,J_2 \ldots$. They then show that, firstly, quantum sigma-model
corrections are suppressed by powers of the inverse total charge 
$J=J_1+J_2+ \ldots$, and that, secondly, the obtained expressions
for the string energies again expand in {\it integer} powers of 
the BMN coupling constant eq.(\ref{eq:lambda'}). The latter
fact led Frolov and Tseytlin to propose that, just as in the BMN case,
the conjectured dual gauge theory operators should possess
anomalous dimensions whose perturbative expansions match the just
mentioned string energy expansions.
An added benefit of this suggestion is that the symmetry charges
$J_i$ allow for a simple identification of those conformal ${\cal N}=4$
scaling operators which are natural candidates for the gauge theory
analogs of the string theory states. The proposal was further
elaborated on a large number of explicit examples 
\cite{ftfollowup},\cite{ft4}. In all cases the classical problem can
be exactly solved, the underlying reason being, once again,
integrability \cite{afrt} (classical, however, in this case).

The final (all-orders in $\lambda'$) expressions for the 
classical string energies are rather non-trivial functions of the 
charge ratios $J_i/J$, as they are obtained by solving a non-linear 
system of equations. This is even true for the leading
${\cal O}(\lambda')$ term, which should correspond to a one-loop
anomalous dimension in gauge theory. By applying the 
above mentioned Bethe ansatz to the relevant operators, it was 
indeed shown in a number of non-trivial examples that the leading string
predictions can be reproduced by one-loop ${\cal N}=4$ gauge theory
\cite{bmsz},\cite{bfst},\cite{emz}. The relevant quantum spin chain
is integrable since it possesses an infinite number of mutually
commuting charges. As was recently shown in \cite{as},
these may be reproduced from the Bethe ansatz
solutions as well, and precisely {\it match} the corresponding infinite
tower of string sigma-model charges to leading order in $\lambda'$.
For a recent, up-to-date review of ``spinning strings'', see \cite{arkady}.

The last result \cite{as} would seem to be a near-proof, assuming the
correctness of AdS/CFT, that (1) ${\cal N}=4$ is integrable to 
{\it all} orders in $\lambda'$,
and that (2) the BMN and Frolov-Tseytlin (FT) proposals 
should indeed be valid to {\it all} orders in 
perturbation theory. The charges certainly are  
commuting for all values of $\lambda'$ on the string side,
and it would appear to be very difficult to imagine that there
exist two inequivalent deformations of the matching leading
order ${\cal O}(\lambda')$ commuting charges. Some caution is
nevertheless warranted. Curious, unexpected and unaccounted
structural agreements
between small $\lambda$ gauge theory and large $\lambda$ string
theory results have previously appeared in the context of AdS/CFT, see
in particular \cite{bianchi-gleb}. One should definitely push
perturbative ${\cal N}=4$ to higher orders and see whether the 
agreement with BMN/FT persists.

The BMN prediction, originally checked in
\cite{bmn} at the one-loop level, was successfully tested in
\cite{gross},\cite{eden} at two loops. 
It is usually stated in the recent literature
that in \cite{sz} an all-orders proof was given. 
It appears to be difficult to rigorously justify some of the details 
of the proof, and we remain agnostic with regards to its validity. 
At any rate, while inspiring, the methodology of \cite{sz} is not of a 
constructive nature and we are unsure whether it does not, just like 
string theory, implicitly assume large $\lambda$.

In \cite{bes3} a program was begun to derive the ${\cal N}=4$
dilatation operator beyond the one-loop level. The two-loop
correction for two complex scalars was found, and interpreted as a 
next-nearest neighbor deformation of the nearest neighbor Heisenberg 
spin chain. Excitingly, the result showed that the two-loop terms
do not lift certain spectral planar one-loop degeneracies,
which could be traced to the integrability property of \cite{mz}.
This led to the conjecture \cite{bes3} that the exact dilatation operator
might correspond to an {\it integrable long-range spin chain}.
Assuming integrability, and BMN scaling, the three-loop correction 
was derived. Using its explicit form, a numerical two- and three-loop
gauge theory estimate for the FT prediction for the 
``folded string'' \cite{ft4}
(see chapter 3 below) was obtained in \cite{bfst}. It showed excellent 
agreement at two loops, but a strange 17\% deviation at three loops.
As a second sign of potential three-loop trouble, a string calculation of the
leading $1/J$ correction to the BMN anomalous dimension formula
was performed in \cite{callan} and, while beautifully agreeing
at two loops, it {\it dis}agreed at three loops with \cite{bes3}.

Important further information comes from recent work by 
Beisert \cite{dynamic}. The method goes beyond \cite{bes3} and
uses, apart from field theoretic structural properties,
symmetry arguments in order to constrain
the three-loop dilatation operator (actually, for a larger
class of fields than just two complex scalars). For our
present purposes (we stick to the latter case in the present
paper), the upshot of his study is as follows:
(1) Perturbative integrability (in the sense of \cite{bes3})
extends to at least three loops. (2) The two-loop dilatation operator
is fixed up to one constant. This constant may be fitted to
the known two-loop anomalous dimension of the Konishi field. This
way one independently verifies the two-loop BMN prediction, in 
agreement with \cite{gross},\cite{eden}. 
(3) The three-loop dilatation operator
is determined up to two unknown constants. These could be found if
we knew two (independent) ${\cal N}=4$ anomalous dimensions
(such as Konishi, plus one further field). However, no three-loop
${\cal N}=4$ anomalous dimensions are rigorously known to
date\footnote{
It would clearly be of great importance to perform a reliable
field theoretic computation of the planar three-loop anomalous dimensions
of two independent fields in ${\cal N}=4$ gauge theory.}.
Therefore, the three-loop dilatation operator is not yet 
rigorously known. However, {\it if} we impose qualitative
BMN scaling ({\it i.e.}~if we assume in accordance with
eq.(\ref{eq:lambda'}) that the ${\cal O}(\lambda^3)$ anomalous
dimensions of finite $J$ BMN operators \cite{beisert1} scale like
$\sim J^{-6}$), both constants are fixed and the infinite $J$ BMN result is
also quantitatively reproduced\footnote{However, as we just mentioned,
the $1/J$ corrections are {\it not} properly reproduced \cite{callan}.}.

In an interesting parallel development, it was established that
a system closely related to full-fledged ${\cal N}=4$ gauge
theory, plane-wave matrix theory, also exhibits integrability to
at least three loops \cite{jan}. This is a quantum mechanical
system, and this result was rigorously derived without further
assumptions.

Summarizing the last paragraphs, integrability seems to be a very
stable concept in ${\cal N}=4$ gauge theory even beyond one-loop;
with further, indirect evidence coming from the string side of
the AdS/CFT correspondence \cite{wadia},\cite{afrt},\cite{as}.
It is therefore important to gain deeper understanding of 
the integrable long range spin chain potentially describing
planar ${\cal N}=4$ gauge theory {\it non-perturbatively}.
Such a spin chain, in contrast to the Heisenberg model,
should contain an extra parameter, related to the Yang-Mills
coupling constant. There exists one well-known integrable long-range
spin chain, namely  the Haldane-Shastry chain \cite{haldaneshastry}.
However, there the parameter corresponds to the length of the
chain\footnote{This is reminiscent of the
BMN coupling constant eq.(\ref{eq:lambda'}), which also depends
on the length of the spin chain. From our point of view,
this is not an accident.}. In particular, it is not possible to
recover the one-loop Heisenberg model from Haldane-Shastry.
There also exists a much less known integrable spin chain that 
precisely furnishes such an extra deformation parameter,
the Inozemtsev long range spin chain \cite{inozemtsev}.
Interestingly, it encompasses both the Haldane-Shastry chain and
the Heisenberg chain. 
It is the purpose of this paper to investigate whether it
might serve as a candidate for describing the sought all-loop 
dilatation operator (for two complex scalars). We shall find that
this is indeed consistent with all currently known solid information.
However, proving or disproving it will require more work.

The paper is organized as follows. In chapter two we describe
the Inozemtsev spin chain, and show that it is indeed capable
of emulating the ${\cal N}=4$ dilatation operator up to at least three loops.
It does not fix the yet to be determined two constants of the
three-loop terms, as discussed above. This is not surprising,
as they do not affect integrability \cite{dynamic}. We also describe a Bethe ansatz
for this quantum chain, which is due to Inozemtsev as well (see
\cite{ino02}, and references therein). This ansatz (unlike
the full Inozemtsev spin chain) is limited to operators
which are ``long'', which means that the interaction range is
kept smaller than the length of the chain. We employ it to
re-derive all known results \cite{bes3} for finite $J$ BMN operators.
The Inozemtsev-Bethe ansatz is particularly interesting
since it may, in principle, be used to derive non-perturbative 
results. In chapter three we apply the ansatz to the
case of spinning strings. We are able to derive some of the
highly involved {\it two-loop} expressions of \cite{ft2,ft4,afrt} for
folded and circular strings, in agreement with our previous
numerical study for a particular case \cite{bfst}. However,
at three loops we find that gauge theory yields a similar,
but {\it different} result as compared to the FT proposal. 
On the positive side, we very accurately reproduce the previous numerical
estimate for a special case \cite{bfst}.

Is this really bad news? Not necessarily. Of course, the perspicacious
reader  might suspect that the three-loop dilatation operator
as conjectured in \cite{bes3} and derived in \cite{dynamic} is simply wrong.
In the absence of an independent, rigorous field theory computation,
this remains a theoretical possibility; however, we currently
do not see which mistake could have been made. On the other hand,
it might be the case that the BMN and FT results are only
valid at large $\lambda$ (the one- and two-loop agreement might
be due to some yet-to-be understood planar weak-strong coupling 
duality symmetry); the expansion is, after all, in $\lambda'$,
not $\lambda$. This issue came up already in certain puzzles
related to the BMN limit \cite{ksv}.
In this context it is interesting that the 
Inozemtsev chain exhibits a generic breakdown of {\it perturbative}
BMN scaling at {\it four} loops\footnote{
We note in passing that we suspect
the Inozemtsev model to be equally capable of describing the 
plane-wave matrix model mentioned above,
which has also been shown to be ``perturbatively'' integrable to at
least three loops \cite{jan}. There BMN scaling
may be recovered, up to three loops, by a redefinition of the mass
parameter of the model. If our Inozemtsev ansatz is correct
we suspect that such a redefinition is no longer possible at
the four-loop level (or else, integrability of the model should break
down). This might be a more feasible check of 
our Inozemtsev ansatz as opposed to considering four-loop perturbation
theory in full-fledged ${\cal N}=4$ field theory. In turn, if the
plane wave mechanics stays integrable at four loops while still being
consistent, after redefinition, with BMN scaling, it would strongly
suggest that there exist further integrable long range spin chains
differing from Inozemtsev.  }. However, it is possible
to recover BMN scaling non-perturbatively, {\it i.e.}~at strong 
coupling (see chapter 2). 
A further, more troubling explanation for the deviation
we are finding would be a large $N$ phase transition when we go from weak to strong coupling \cite{mike}.

In this paper we have shown how to extend the Bethe ansatz
for planar one-loop ${\cal N}=4$ gauge theory to higher loops.
Our procedure is completely sound up to three loops.
Of course we are far from proving that our proposal is valid beyond 
three loops, let alone non-perturbatively.  However,
having identified an integrable long-range spin chain
which properly describes the dilatation operator up to three loops,
we should take it seriously. Integrability is a 
subtle and tight structure, and not found abundantly.

We have not been able to explicitly mention in this introduction
many interesting, closely related recent contributions to this
subject of study, see in particular \cite{further}.

\section{The Inozemtsev spin chain and ${\cal N}=4$ gauge theory}

\subsection{Long-range spin chains with elliptic 
and hyperbolic interactions}

More than one decade ago, Inozemtsev \cite{inozemtsev} 
proposed a spin chain with long range 
interaction which interpolates between the Heisenberg  and 
Haldane-Shastry \cite{haldaneshastry} spin chains. His Hamiltonian takes the form
\begin{equation}
\label{ino}
I=\sum_{j=1}^L  \sum_{n=1} ^{L-1} {\cal P}_{L,\pi/\kappa}(n) (1-P_{j,j+n})\;,
\end{equation}
where the operator $P_{jk}$ permutes the spins at  sites $i$ and $j$,
and the interaction strength is defined in terms of
the elliptic Weierstrass function ${\cal P}_{L,\pi/\kappa}(n)$
with periods $L$ and $i \pi/\kappa$. We recall that the Weierstrass function  can be defined by the following series
$${\cal P}_{L,\pi/\kappa}(z)=\frac{1}{ z^2} +\sum'_{m,n}\left( 
\frac{1}{ (z-mL-i n\pi/\kappa)^2}-
\frac{1}{ (mL+i n\pi/\kappa)^2}\right)\;,$$
where the prime means that the term with $m=n=0$ is omitted.
When one of the two periods becomes infinite, the Weierstrass function
becomes a trigonometric or hyperbolic function. 
If $\kappa \to 0$, this function goes smoothly to the
Haldane-Shastry interaction,
$$\lim_{\kappa\to 0}{\cal P}_{L,\pi/\kappa}(z)=\left(\frac {\pi }{L}\right)^2
\left( \frac{1}{ \sin^2 \pi z /L}-\frac{1}{ 3}\right)\;,$$
while for $L\to\infty$ the interaction takes the form of a hyperbolic
function, and decays exponentially with the distance,
$$\lim_{L\to \infty}{\cal P}_{L,\pi/\kappa}(z)=\kappa^2
\left(\frac {1}{ \sinh^2 \kappa z }+\frac{1}{ 3}\right)\;.$$

The Heisenberg limit can be recovered in the limit $\kappa \to \infty$,
where the imaginary period vanishes.
To be able to properly take this limit we define the ``coupling
constant'' 
\begin{equation}
\label{eq:tdef}
t \equiv e^{-2\kappa}\;, 
\end{equation}
and we  use a redefined interaction strength,
by removing an additive and a multiplicative 
constant 
$$  h_{t,L}(n) \equiv \frac{{\cal P}_{L,\pi/\kappa}(n)}{4\kappa^2}-\frac{1}{12}=
\sum_{m=-\infty}^{\infty}\frac {t^{n+Lm}}{ (t^n-t^{Lm})^2}\;.$$
The Heisenberg interaction can be obtained to leading order in $t$
of the Inozemtsev interaction 
$$\lim_{t \to 0}
h_{t,L}(n)/t=\delta_{n,1}+\delta_{n,L-1}\;.$$
Going beyond the first order in $t$, we notice that $h_{t,L}(n)$ starts
as $t^n$ (plus higher orders in $t$), as long as $n<L/2$, therefore the Inozemtsev interaction involves spins separated by a distance of at most $n$ at order $n$ in $t$.  
This property agrees with the structure of the perturbative dilatation operator
in the planar limit \cite{gross}.

Inozemtsev performed a detailed study of the Hamiltonian
(\ref{ino}) (see  \cite{ino02}  for a review) and gave convincing arguments
in the favor of its integrability, by finding the corresponding Lax pair
\footnote{According to Inozemtsev \cite{ino02}, who invokes a result by Krichever
\cite{krichever},
this is the most general Lax pair corresponding to a spin chain with
long range interaction and pairwise exchange given by permutation
operator. 
However, his arguments do not directly apply to the case when the
interaction involves more than two spins, and we should keep an
open mind as far as concerns the existence of integrable long
range spin chains of a more general type than Inozemtsev.
For these putative chains, the subtraction of higher charges from the
lowest charge ({\it i.e.}~the Hamiltonian) would not allow one to
eliminate interactions involving more than two spins.
}. 
In addition, he explicitly constructed the first of the higher
charges commuting with the Hamiltonian. The existence of this first
higher charge usually indicates complete integrability for spin chains.
However, the general form of the
commuting charges or the monodromy matrix are still out of reach.
The connection with the Calogero-Moser Hamiltonian  with elliptic potential
presented in \cite{ino02} may help finding the exact solution of this chain.
Among the aforementioned
limiting cases, the Heisenberg case $\kappa\to \infty$ and 
the Haldane-Shastry case, $\kappa\to 0$ are completely solved, the first using the
Bethe ansatz and the second by exploiting the Yangian symmetry \cite{HS}.

In the following, we are going to concentrate on the hyperbolic case,
where the length of the chain becomes infinite ({\it i.e.}~$L\to \infty$), 
and the interaction takes the simpler form 
$$  h_t(n) =\frac{1}{4\sinh^2 \kappa n}= \frac{t^n}{ (t^n-1)^2}\;,$$
that is, the periodicity effects of the interaction are negligible.
This case is easier to treat analytically, but still quite interesting from 
the point of view of comparison with the known results for the 
${\cal N}=4 $ SYM theory. 
In particular, it allows to study 
BMN operators at finite (but large) $J\simeq L$. 
The full elliptic potential takes into account the effects related to the periodicity
of the chain, therefore one could hope that it contains the information concerning ``short'' operators such as Konishi.

In the hyperbolic regime we can use an analytic continuation for the monodromy
matrix and for the conserved charges of the Haldane-Shastry spin chain
\cite{HS,BGHP,HT}. 
The Haldane-Shastry conserved charges
still commute (and commute with the Yangian generators) if one replaces the coordinates $z_k=e^{2\pi i k/L}$
by $z_k=t^{-k}$ \cite{BGHP}
 \footnote{The generating function for the conserved 
charges of the Haldane-Shastry Hamiltonian was identified
in \cite{HT} by taking the limit $\lambda\to \infty$ of the monodromy matrix for the Calogero-Sutherland model with spin. The same argument
can be easily applied to the hyperbolic case, 
since in this case the leading term of the quantum determinant is
trivial, 
$\Delta_0(u)=0$, as well as its derivatives.
It follows that the next-to-leading term  $\Delta_1(u)$ in
$\lambda^{-1}$
in the expansion of the quantum determinant can be used to 
generate the commuting charges in exactly the 
same fashion as for Haldane-Shastry spin chain.
}.
 We recall that the Yangian generators
are given by \cite{HS}
\begin{eqnarray}
Q_0^a&=&\sum_j\sigma^a_j\;,\cr
Q_1^a&=&i\varepsilon^{abc}\sum_{j<k}\frac{z_j+z_k}{z_j-z_k}\;
\sigma^b_j\,\sigma^c_k\;.
\end{eqnarray}
Let us emphasize that the Yangian is a true symmetry ({\it i.e.}~it 
commutes with the conserved charges) only for the hyperbolic chain,
when the length of the chain is infinite. Otherwise, like in the case 
of Heisenberg spin chain \cite{denis}, 
in the full elliptic case the Yangian symmetry is broken by ``boundary terms''.
The higher conserved 
charges of the hyperbolic spin model are also obtained 
by simply replacing $z_k=t^{-k}$ in the expression ofor the 
Haldane-Shastry conserved
charges
\cite{HS,HT}
\begin{eqnarray}
 \label{conserved}
I_2&=&\sum'_{ij}\frac{z_iz_j}{z_{ij}z_{ji}}(P_{ij}-1) \;,\cr
I_3&=&\sum'_{ijk}\frac{z_iz_jz_k}{z_{ij}z_{jk}z_{ki}}(P_{ijk}-1)\;,\cr
I_4&=&\sum'_{ijkl}\frac{z_iz_jz_kz_l}{z_{ij}z_{jk}z_{kl}z_{li}}(P_{ijkl}-1)-
 2\sum'_{ij}\left(\frac{z_iz_j}{z_{ij}z_{ji}}\right)^2(P_{ij}-1) \;,
\end{eqnarray}
where the prime indicates that the sum is over distinct summation
indices, $z_{ij}=z_i-z_j$, and  $P_{ijk}$, etc. represent cyclic
permutations. A systematic procedure to find all the conserved charges
was given in 
\cite{HT}, although the explicit construction of these charges 
becomes  more and more involved at higher orders.

The diagonalization of the conserved charges is achieved, 
in the case of nearest-neighbor interaction, by the algebraic Bethe ansatz.
In the case of the long-range interaction, the algebraic Bethe ansatz does
not work any more, just like in the case of the Haldane-Shastry spin chain.
There, the diagonalization can be performed by using the 
Yangian symmetry and the relation to the 
Calogero-Sutherland model 
\cite{HS,Cargese}.
In the hyperbolic case
it is still possible
to use the so-called asymptotic Bethe ansatz,
valid in the regime
where $L$
is very large but still
finite. The phase acquired by a quasi-particle (magnon) traveling around the 
circumference of the chain is given by the phase gained by its scattering  with all the other magnons in the system
\begin{equation}
\label{ABA}
\exp{(ip_jL)}=\exp\left(i\sum_{\genfrac{}{}{0pt}{}{k=1}{k\neq j}}^M
\chi(p_j,p_k)\right)\;,
\end{equation}
where $\chi(p_j,p_k)$
is the phase shift for the scattering 
of two magnons of momenta
$p_j$ and $p_k$ and $M$ is the number of magnons.
Inozemtsev computed the phase shift for the
scattering of two
magnons, 
\begin{eqnarray}
        \label{phaseshift}
&\ &\cot  \frac{\chi(p_j,p_k)}{2}=\varphi(p_j)-\varphi(p_k)\;,\cr
&\ & \cr
&\ &\varphi(p)=\frac{p}{2\pi i \kappa}\zeta_1\left(\frac{i\pi}{2\kappa}
\right)-\frac{1}{2 i \kappa}\zeta_1\left(\frac{i p}{2\kappa}
\right)\;,
\end{eqnarray}
as well as the magnon energy,
\begin{equation}
\label{magnen}
\varepsilon(p)=
\frac{1}{i\pi\kappa}\zeta_1\left(\frac{i\pi}{2\kappa}
\right)-\frac{1}{4\kappa^2}{\cal P}_1\left(\frac{i p}{2\kappa}
\right)-\varphi^2(p)\;,
\end{equation}
where ${\cal P}_1(z)={\rm d} \zeta_1(z)/{\rm d}z$ denotes the Weierstrass elliptic functions 
of periods $1$ and $i\pi/\kappa$, and the quasi-periodic 
zeta function is odd, $\zeta_1(-z)=-\zeta_1(z)$.
The function $\varphi(p)$ is also quasi-periodic, $\varphi(p+2\pi)=
\varphi(p)$ and $\varphi(p+2i\kappa)=\varphi(p)-i$.
While the phase shift and magnon energy  (\ref{phaseshift},\ref{magnen}) are
exact, the Bethe ansatz formula (\ref{ABA}) is asymptotic,
{\it i.e.}~it is valid for chains of finite but large length $L$.
It is easy to check  that in the  limit  $\kappa \to\infty$,
the magnon energy and the phase shift reduce to the corresponding Heisenberg 
values, in particular $\varphi(p)\to (1/2) \cot (p/2)$.

\subsection{Comparing the conserved charges of the Inozemtsev spin
chain to the dilatation operator}

If we assume that the dilatation operator in the $su(2)$ 
sub-sector of ${\cal N}=4$ theory
is integrable to all orders in perturbation theory, 
as first conjectured in \cite{bes3}, it should correspond to some
integrable long-range spin chain. We believe that 
the Inozemtsev spin chain is an interesting and
natural candidate. As we shall prove, it is definitely capable of
describing the dilatation operator in the subsector to at least 
three loops. However, if the describtion extends to four loops
and beyond, perturbative BMN scaling would have to be violated 
at the four loop level.

If the dilatation operator contains, beyond two loops,
interaction terms which involve more than two spins,
the Inozemtsev Hamiltonian is not, by itself, sufficient to reproduce 
the former\footnote{The three-loop dilatation operator
proposed in \cite{bes3} does contain four-spin terms, as these
are necessary for perturbative BMN scaling. If we drop this
requirement, the yet undetermined two constants \cite{dynamic} 
discussed in the introduction may
be chosen so as to eliminate the four spin terms. The future will
tell.}. In this case, one should consider a linear combination of the
even parity
conserved charges of the Inozemtsev spin chain\footnote{In \cite{bes3}
it was argued that only 
connected planar diagrams contribute to the dilatation operator
(see also \cite{gross}). A completely rigorous all-orders proof
is not easy due to the known problems with regularization by 
dimensional reduction
at large loop order in ${\cal N}=4$ gauge theory. Non-connected
diagrams would require 
non-linear combinations of the conserved charges.}. The most general
expression for the dilatation operator will then be  
\begin{equation}
\label{dilat}
D(\lambda)= L+f_1(\lambda)I_2(t)+f_2(\lambda)I_4(t)+f_3(\lambda)I_6(t)+...
\end{equation} 
where $t=t(\lambda)$ is such that $\lim_{\lambda\to 0}t/\lambda=1/16\pi^2$,
$f_k(\lambda)$ are functions 
regular at $\lambda=0$ and 
$I_{2k}(t)$ are the even parity
conserved charges of Inozemtsev spin chain.

To determine the  precise relation between the 't Hooft coupling constant $\lambda$
 and the parameter $t$, as well as the functions
$f_k(\lambda)$, we have to use arguments going beyond 
perturbation theory.
The first example of  such an argument concerns the behavior 
of the rapidity $\varphi(p)$ in the BMN limit, where the  coupling constant $\lambda$ and the chain length $L$ are both large, 
$\lambda=\lambda'\,L^2$. 
Since it enters the Bethe ansatz equations, and therefore
all the  observables, the combination $\varphi(p)/L$ should have a well-behaved BMN limit.
The rapidity (\ref{phaseshift}) can be written as
an infinite sum
\begin{eqnarray}
\label{uphi}
\varphi(p)&=&\frac{1}{2}\cot\frac{p}{2}+\frac{1}{2}
\sum_{n>0}\left[\cot(\frac{p}{2}-i\kappa n)+\cot(\frac{p}{2}+i\kappa n)
\right] \\
&=&\frac{1}{2}\cot\frac{p}{2}  
+2\sum_{n>0}\frac{t^n \sin p}{(1-t^{n})^2+4t^n\sin^2 (p/2)}\;. \nonumber
\end{eqnarray}
where in the second line we have used the perturbative parameter $t$ 
instead of the parameter $\kappa$. Expanding this result to the first few orders 
in $p$ we obtain 
$$
\varphi(p)=
\frac{1}{2}\cot\frac{p}{2}+2p\sum_{n>0}\frac{t^n}{(1-t^n)^2}-
\frac{p^3}{3} \sum_{n>0} \frac{t^n(1+4 t^n+t^{2 n})}{(1-t^n)^4}
+\ldots\;.
$$
In the BMN limit, the leading behavior of the magnon momentum 
is $p\sim 1/L$, and therefore the rapidity scales as
$\varphi \sim 1/p \sim L$. 
In order to obtain a finite but non-zero result for the rescaled rapidity
$\varphi /L$
in the perturbative regime $\lambda \sim 0$ where $\kappa \to \infty$
(i.e.~$t \to 0$) we see that 
we need to identify
\begin{equation}
\label{tlambda}
\frac{\lambda}{16\pi^2}=\sum_{n>0}\frac{t^n}{(1-t^n)^2}
=\sum_{n>0}\frac{1}{4\sinh^2(\kappa n)}\;.
\end{equation}
On then finds that
\begin{eqnarray}
\label{breakdown}
\frac{\varphi}{L}&=&\frac{1}{L p}+\frac{2}{L^2} \frac{\lambda}{16 \pi^2} (L p)-
\frac{1}{3 L^4}\left[\frac{\lambda}{16 \pi^2}
+6 \left(\frac{\lambda}{16 \pi^2}\right)^2
-12 \left( \frac{\lambda}{16 \pi^2} \right)^3 \right] (L p)^3+\ldots \cr
 &=& \frac{1}{L p}+ \frac{\lambda'}{8 \pi^2}  (L p) 
-\frac{\lambda'^2}{128 \pi^4} (L p)^3
+L^2~\frac{\lambda'^3}{1024 \pi^6} (L p)^3+\ldots
\qquad {\rm as} \qquad L \to \infty
\end{eqnarray}
where we have used the finite BMN coupling constant $\lambda'=\lambda/L^2$.

We see that at ${\cal O}(\lambda'^3)$ a quadratic divergence appears
in the perturbative expression for the rescaled rapidity $\varphi /L$.
The consequences of this perturbative scaling violation will be
discussed in section 2.4 below. 
 
It is interesting to see that a finite rescaled rapidity $\varphi /L$
may be obtained {\it provided} we stay in the strong-coupling regime
where $\kappa\to 0$ (i.e.~$t \to1$) and thus 
$\lambda=2\pi^4/3\kappa^2 \to \infty$.
Here, it is useful to use a dual representation for the rapidity $\varphi(p)$
\begin{equation}
\label{asymptphase}
\varphi(p)=\frac{\pi}{2\kappa}\coth\frac{\pi p}{2\kappa}+
\frac{\pi}{2\kappa}\sum_{m>0}\left[\coth\frac{\pi(p-2\pi m)}{2\kappa}+
\coth\frac{\pi(p+2\pi m)}{2\kappa}
\right]-\frac{p}{2\kappa}\;,
\end{equation}
where the last term insures the desired quasi-periodicity properties, 
$\varphi(p+2\pi)=\varphi(p)$ and $\varphi(p+2i\kappa)=\varphi(p)-i$.
The terms  with $m>0$ form a power series
in the dual parameter $e^{-2\pi^2/\kappa}$ and  they vanish 
exponentially in the  BMN limit, since in this case $1/\kappa\sim 
L\to\infty$. The remaining part obviously obeys  BMN scaling
and may be written as
\begin{equation}
\label{coth}
\varphi(p)=\frac{1}{2 \pi} \sqrt{\frac{3}{2} \lambda'} 
\coth \frac{p}{2 \pi} \sqrt{\frac{3}{2} \lambda'}\, ,
\end{equation}
where we have replaced $\varphi(p)/L$ by $\varphi(p)$ and $p L$ by
$p$.

In principle, the explicit knowledge of all the conserved 
charges of the Inozemtsev chain and of their eigenvalues
would allow to obtain information about the functions 
$f_k(\lambda)$ in equation (\ref{dilat}). Since, at least for the moment, we lack this knowledge we have to restrict ourselves to comparing the relation (\ref{dilat}) to the perturbative results which are already known
 \cite{bes3},\cite{dynamic}. 
Let us then compare the expression (\ref{dilat}) to the dilatation operator,
$$D(\lambda)=L+\sum_{k>0}\left(\frac{\lambda}{16\pi^2}\right)^kD_{2k}$$
as stated in \cite{bes3}.
Rewritten using commuting permutations, the contributions to the dilatation operator up to three loops are
\begin{eqnarray}
\label{pertdilat}
D_0&=&L\;,\\  \nonumber
D_2&=&2L-2\sum_i P_{i,i+1}\;,\\ \nonumber
D_4&=&-6L+8\sum_i P_{i,i+1}-2 \sum_i P_{i,i+2}\;,\\ \nonumber
D_6&=&40L-56\sum_i P_{i,i+1}+16\sum_i P_{i,i+2}
-4\sum_i \left(P_{i,i+3}P_{i+1,i+2}- P_{i,i+2}P_{i+1,i+3}\right)\;.
\end{eqnarray}
We recall that  $D_2$ was  obtained by diagrammatic computation
in ${\cal N}=4$ SYM theory and the expression of $D_4$
was checked in several ways \cite{gross},\cite{eden}.
The expressions for $D_6$  were obtained initially 
\cite{bes3,bes4}
supposing perturbative integrability and imposing perturbative BMN
scaling. 
The expression for $D_6$ was subject to controversies,
since it disagrees with the results from the near plane-wave string
theory \cite{callan}.
Later on, Beisert \cite{dynamic}  checked that the 
integrability of $D_6$ is insured by supersymmetry, and that the 
two free coefficients are fixed if one imposes perturbative BMN scaling.

The conserved charges of the Inozemtsev spin chain can be 
expanded in series of $t$, and therefore 
of the coupling constant $\lambda$. 
By inverting the series (\ref{tlambda}) we 
obtain 
\begin{equation}
\label{lambdat}
t=\frac{\lambda}{16\pi^2}-3\left(\frac{\lambda}{16\pi^2}\right)^2+
14\left(\frac{\lambda}{16\pi^2}\right)^3+\ldots,
\end{equation} 
Inserting this expression into (\ref{conserved}) and expanding in powers of $\lambda$ we get
$$I_{2k}=\sum_{n>0}\left(\frac{\lambda}{16\pi^2}\right)^n I_{2k}^{(n)}\;.
$$
The explicit expression for the first few orders in the Hamiltonian is
\begin{eqnarray*}
I_{2}^{(1)}&=&2L-2\sum_i P_{i,i+1}\;,\\
I_{2}^{(2)}&=&2\sum_i P_{i,i+1}-2\sum_i P_{i,i+2}\;,\\
I_{2}^{(3)}&=&-10\sum_i P_{i,i+1}+12\sum_i P_{i,i+2}-2\sum_i P_{i,i+3}\;.
\end{eqnarray*}
In the following, we use a redefinition of the fourth conserved
charge, $\tilde I_4(\lambda)=I_4(\lambda)-\frac{\lambda}{8 \pi^2} I_2(\lambda)$.
This is done for the purpose of having the fourth conserved
charge of the Heisenberg chain as the first non-trivial
order in the $\lambda$ expansion 
\begin{eqnarray*}
\tilde I_{4}^{(1)}&=&\tilde I_{4}^{(2)}=0\;,\\
\tilde I_{4}^{(3)}&=&-8L+8\sum_i P_{i,i+1}-4\sum_i P_{i,i+2}+4\sum_i P_{i,i+3}\\ &-&
8\sum_i \left(P_{i,i+3}P_{i+1,i+2}- P_{i,i+2}P_{i+1,i+3}\right)\;,
\end{eqnarray*}
It is not obvious that a similar redefinition can be done 
for all the conserved charges, but at least we know that we can
arrange to have  $\tilde I_6^{(3)}=0$.

We can now reconstitute the perturbative dilatation operator
(\ref{pertdilat}) using the Inozemtsev conserved charges
\begin{eqnarray}
\label{dilatIno}
\nonumber D_2&=&I_{2}^{(1)}\;,\\ \nonumber
D_4&=&I_{2}^{(2)}-3I_{2}^{(1)}\;,\\
D_6&=&I_{2}^{(3)}+\frac{1}{2}\tilde I_{4}^{(3)}-3I_{2}^{(2)}+22I_{2}^{(1)}\;.
\end{eqnarray}
If we reinstate the constants which were fixed by imposing perturbative BMN scaling \cite{bes3}, the most general form for the third loop dilatation operator allowed by perturbative integrability is
\begin{equation}
\label{nonBMN}
D_6=I_{2}^{(3)}+a_1\tilde I_{4}^{(3)}-3I_{2}^{(2)}+a_2I_{2}^{(1)}\;.
\end{equation}
Comparing to the expression (\ref{dilat}), we conclude that
$f_1(\lambda)=1-(2a_1+3) \frac{\lambda}{16 \pi^2}
+\ldots$ and $f_2(\lambda)=a_1+\ldots$, 
the constant $a_2$ appearing at next order in $\lambda$. 
In conclusion, we can reproduce the dilatation 
operator up to three loops using the conserved charges of 
the Inozemtsev spin chain.

\subsection{Perturbative results from the asymptotic Bethe ansatz.
 Dimensions of the BMN operators}

In the following, we are going to compute perturbatively the dimensions
of the BMN operators up to third order, supposing that the dilatation operator is  integrable and its integrability 
is governed by the Inozemtsev spin chain. The strategy is to compute the magnon momenta by using the asymptotic Bethe ansatz
equation (\ref{ABA}) and to substitute them into the expression of the dilatation operator known 
from perturbative calculations.

The asymptotic Bethe ansatz equations have the form (\ref{ABA}), which
we can rewrite as 
\begin{equation}
\label{eq:ino-bethe}
e^{i p_j L}=\prod_{\genfrac{}{}{0pt}{}{k=1}{k\neq j}}^{M}
\frac{\varphi_j-\varphi_k+i}{\varphi_j-\varphi_k-i}\;,
\end{equation}
where $M$ is the magnon number and the rapidity $\varphi_j=\varphi(p_j)$ is given by the expression (\ref{uphi}).
To see the connection with the usual Bethe ansatz for the 
Heisenberg chain, we can replace the momentum by a rapidity
variable $u_j=\frac{1}{2}\cot\frac{p_j}{2}$ such that the left
hand side of eq.(\ref{eq:ino-bethe}) becomes 
$$
e^{i p_j L}=\left( \frac{u_j+i/2}{u_j-i/2} \right)^L,
$$
while it is clear from eq.(\ref{uphi}) that 
$\varphi_j=u_j+ {\cal O}(t)$. More precisely, up to third order 
in $t$ we have
$$
\varphi(p)=u(p)+2t\sin p+2t^2\sin p\,(1+2\cos p)+8t^3\sin p \cos^2 p
+\ldots\;.
$$

We can 
perturbatively  solve the Bethe ansatz equation for the two-magnon case, with the two magnon momenta satisfying
$p_1=-p_2=p$. 
Since we will compare our results to the anomalous 
dimensions of the BMN operators
$$\mathop{\mathrm{Tr}} Z^{J} \Phi^{2}  + \ldots \ ,$$
 we 
are going to use $J+2$ to denote  the length of the chain,
$$L=J+2\;.$$
The Heisenberg solution for the magnon momentum 
is 
$$p=p_n\equiv\frac{2\pi n}{J+1}\;,$$
and the corrections to this value are given by the equation  
\begin{eqnarray*}
p&=&p_n-\frac{8t\sin p \sin^2 \frac{p}{2}}{J+1}-\frac{8t^2\sin p \sin^2 \frac{p}{2}}{J+1}
\left(8 \cos^4 \frac{p}{2}-4\cos^2 \frac{p}{2}+1\right)\\
&-&\frac {32t^3\sin {p}\,  \sin^2 \frac{p}{2} }{3(J+1)} 
 \left(64 \cos^8 \frac{p}{2}   -96 \cos^6 \frac{p}{2}  +48
 \cos^4 \frac{p}{2}   -16
\cos^2 \frac{p}{2} +3 \right) +\ldots\;.
\end{eqnarray*}
The perturbative solution of the Bethe ansatz equation up to third
order in $t$ is
\begin{eqnarray*}
p&=&p_n-\,{\frac {8t \sin {p_n}\sin^2 \frac{p_n}{2}  
 }{J+1}}\\
& +& \frac{ 64t^2\,
 \sin {p_n} \sin ^4\frac{p_n}{2}     \left( 4 \cos^2  \frac{p_n}{2}
-1 \right)} {{(J+1)}^{2}}
 -\frac{8t^2\sin {p_n}\,  \sin^2 \frac{p_n}{2}  
 \left(8\,  \cos^4 \frac{p_n}{2}   -4\, \cos^2 \frac{p_n}{2}   -1
\right)} {J+1}\\
&-&
\frac{512t^3 \, \sin {p_n} \sin^6 \left( \frac{p_n}{2}   \right)   \left( 24
 \cos^4 \frac{p_n}{2}   -13
\cos^2\frac{p_n}{2}   +1 \right)}{ {(J+1)}^{3}}\\
&+&\frac{128\,t^3
\sin {p_n} \sin^4 \left( \frac{p_n}{2}   \right)  \left( 48 \cos ^6 \frac{p_n}{2} -44\cos ^4\frac{p_n}{2}
 +4 \cos^2 \frac{p_n}{2}  
+1 \right)} {(J+1)^2}\\
&-&\frac {32t^3\sin {p_n}\,  \sin^2 \frac{p_n}{2}  
 \left(64 \cos^8 \frac{p_n}{2}   -96 \cos^6 \frac{p_n}{2}  +48
 \cos^4 \frac{p_n}{2}   -16
\cos^2 \frac{p_n}{2} +3 \right) }{3(J+1)}+\ldots\;.
\end{eqnarray*}
Using the perturbative expansion (\ref{lambdat}) for $t$ in powers of $\lambda$, 
we can transform this expression into a series in $\lambda$. Then, we insert the solution for the 
momentum in the eigenvalue formula for the dilatation operator, which
is a consequence of (\ref{pertdilat}),
$$E(\lambda)=J+2 \left(1+\frac{\lambda}{2\pi^2} \sin^2\frac{p}{2}
-\frac{\lambda^2}{8\pi^4} \sin^4\frac{p}{2}
+\frac{\lambda^3}{16\pi^6} \sin^6\frac{p}{2}
+\ldots\right)\;.
$$
Using {\tt Maple} to perform the resulting series expansion, we checked
that the  two-magnon energy at order $k$ in $\lambda$, $E^{(k)}$, 
has the behavior 
predicted in \cite{bes3}, up to order $k=3$
\begin{eqnarray}
E^{(1)}&=&\frac{1}{\pi^{2}}\sin^{2}\frac{p_n}{2}\;, \\ \nonumber
E^{(2)}&=&-\frac{1}{\pi^{4}}\sin^{4}\frac{p_n}{2}\left(\frac{1}{4}+
\frac{1}{J+1}\cos^2\frac{p_n}{2} \right)\;,  \\  \nonumber
E^{(3)}&=&\frac{1}{\pi^{6}}\sin^{6}\frac{p_n}{2}\left(\frac{1}{8}+
\frac{\frac{1}{2} \cos^4\frac{p_n}{2}+\frac{3}{4}\cos^2\frac{p_n}{2}}{J+1}+\frac{\frac{5}{2} \cos^4\frac{p_n}{2}-\frac{3}{4}\cos^2\frac{p_n}{2}}{(J+1)^2}
\right)\;.
\end{eqnarray}
The agreement with the perturbative results in gauge theory is  perfect, up to three loops, as far as the anomalous  dimensions of the large $J$ BMN operators are concerned. For completeness, let us look at one more order and, in particular, to compare with the conjectured results of \cite{bes4}.

\subsection{Fourth order dilatation operator and the breakdown of 
perturbative BMN scaling}

The four loop dilatation of ${\cal N}=4$ gauge theory is not yet rigorously
 known. In \cite{bes4} a proposal 
for  its structure was put forward, based (1) on extending
the planar pair symmetry, due to integrability \cite{bes3}, and
(2) on insisting on {\it perturbative} BMN scaling. As we shall see below,
this proposal will turn out to be inconsistent with the Inozemtsev
integrable structure. 
We suspect that the constraints coming from all-order integrability 
become tighter at higher orders  and that it will become 
impossible to satisfy
at once integrability and perturbative BMN scaling. 
However, this does not mean automatically that \cite{bes4}
is wrong. Indeed, it is not excluded that several integrable 
long-range spin chains exist (see the comments in footnotes 4 and 5).

The proposal in 
\cite{bes4} can be rewritten as
\begin{eqnarray}
\label{d8}
D_8&=&222L-648\sum_i P_{i,i+1}+\frac{1340}{3}\sum_i P_{i,i+2}
-\frac{1796}{3}\sum_i P_{i,i+3}+6\sum_i P_{i,i+4}\\ \nonumber
&+&
\frac{842}{3}\sum_i P_{i,i+1}P_{i+2,i+3}+\frac{1972}{3}
\sum_i P_{i,i+3}P_{i+1,i+2}-366\sum_i P_{i,i+2}P_{i+1,i+3}\\ \nonumber
&+&\frac{16}{3}\sum_i( P_{i,i+3}P_{i+2,i+4}+P_{i,i+2}P_{i+1,i+4}-
P_{i,i+4}P_{i+2,i+3}-P_{i,i+4}P_{i+1,i+2}\\ \nonumber
&-&P_{i,i+4}P_{i+1,i+3}+
P_{i,i+3}P_{i+1,i+4})\;,
\end{eqnarray}
where we have chosen the value $\beta=-428/3$ in the expression of \cite{bes4}, since this value allows to reconstitute the 
term in parenthesis, which is a very natural combination of permutation. We recall that $\beta$ corresponds
in \cite{bes4} to a unitary transformation of $D_8$ and that it does not affect the eigenvalue of $D_8$.

We also expand the first two even Inozemtsev conserved charges to fourth order in $\lambda$,
\begin{eqnarray*}
I_{2}^{(4)}&=&62\sum_i P_{i,i+1}-78\sum_i P_{i,i+2}+18\sum_i P_{i,i+3}-2\sum_i P_{i,i+4}\;,\\
\tilde I_{4}^{(4)}&=&48L-36\sum_i P_{i,i+1}+16\sum_i P_{i,i+2}-40\sum_i P_{i,i+3}+12\sum_i P_{i,i+4}\\
&-&8\sum_i P_{i,i+1}P_{i+2,i+3}+
56\sum_i P_{i,i+3}P_{i+1,i+2}- 48\sum_iP_{i,i+2}P_{i+1,i+3}\\
&+&8\sum_i( P_{i,i+3}P_{i+2,i+4}+P_{i,i+2}P_{i+1,i+4}-
P_{i,i+4}P_{i+2,i+3}-P_{i,i+4}P_{i+1,i+2}\\
&-&P_{i,i+4}P_{i+1,i+3}+
P_{i,i+3}P_{i+1,i+4})\;.
\end{eqnarray*}
We see that ansatz (\ref{d8}) can be written as
\begin{equation}
\label{d8I}
D_8=I_{2}^{(4)}+\frac{2}{3}\tilde I_{4}^{(4)}+{\rm lower\ order\ terms}\;,
\end{equation}
where  ``lower order terms''  stands for a term of the form $\sum_i (P_{i,i+1}P_{i+2,i+3}+ P_{i,i+3}P_{i+1,i+2})$ which we expect to
come from
$\tilde I_6^{(4)}$, plus a  linear combination of $\tilde I_{4}^{(3)}$, $I_{2}^{(3)}$,
$I_{2}^{(2)}$, $I_{2}^{(1)}$.

What we learn from this exercise is that, imposing the integrability perturbatively, as it was done in \cite{bes3,bes4}, we obtain the building blocks
of the Inozemtsev spin chain (modulo the contribution
of the sixth conserved charge, for which we do not have a quantitative estimate).
However, the coefficients multiplying $I_{2k}^{(n)}$ in (\ref{dilatIno})  disagree with the all-order expression (\ref{dilat}).
The expression (\ref{dilat}) implies that the coefficient of
$I_{2k}^{(n)}$   in $D_{2n}$ should not depend on $n$. This is not the case
for $I_{4}^{(3)}$ and $I_{4}^{(4)}$ appearing in $D_6$ and $D_8$
in equations (\ref{dilatIno}), (\ref{d8I}).
In order to be compatible with the proposal  (\ref{dilat}), the fourth order dilatation operator has to be
\begin{equation}
\label{newd8}
\tilde D_8=I_2^{(4)}-3 I_2^{(3)}+a_2I_2^{(2)}+b_1I_2^{(1)}+
a_1 \tilde I_4^{(4)}+b_2\tilde I_4^{(3)} +b_3 \tilde I_6^{(4)}\;.
\end{equation}
A possible explanation of the mismatch is related to the fact that the coefficients in $D_6$ and $D_8$ which were left free by perturbative integrability were fixed imposing  perturbative BMN scaling  \cite{bes3,bes4}. 
However, in (\ref{newd8}) there are three free constants left, once we
have fixed, at third order, $a_1$ and $a_2$, while Beisert obtains \cite{bes4}  six constants to be fixed by perturbative BMN. This point  certainly deserves a better understanding.

Independently of the previous analysis, we know that  
the Inozemtsev spin chain cannot reproduce, starting from 
the four-loop order, an anomalous dimension which obeys BMN scaling. 
The reason is our finding in eq.(\ref{breakdown}): 
The rescaled rapidity $\varphi(p) /L$ does not have an 
expansion in $\lambda$ consistent with perturbative BMN
scaling. The first violation appears at 
third order in $\lambda$, and therefore will contribute to the 
correction to the two magnon energy at the {\it fourth order}. 

Although the rapidity (\ref{uphi}) does not obey perturbative BMN scaling (when $\lambda$ is small),  it 
obeys BMN scaling in the BMN regime, when $\lambda$ is 
large. Technically, we see that the perturbative BMN scaling is 
spoiled by terms of the type $e^{-\sqrt {\lambda}}$, 
which vanish in the BMN regime but have a well-defined, non-trivial expansion around $\lambda=0$.

\section{Stringing spins and spinning strings at two and three loops}

\subsection{Generalities}

In the last chapter we showed that the Bethe ansatz for the
Inozemtsev chain is indeed capable of reproducing the correct
two- and three-loop anomalous dimensions of finite length 
two-impurity operators. Let us now turn to the ``thermodynamic''
situation of ``long'' operators with many impurities; {\it i.e.}~we will
study the operators of ``length'' $L=J_1+J_2$ and ``magnon-number''
$M=J_2$
\begin{equation}
\label{eq:embryo}
\mathop{\mathrm{Tr}} Z^{J_1} \Phi^{J_2}  + \ldots \ ,
\end{equation}
in the limit where both $J_1$ and $J_2$ are large. At the one-loop
level anomalous dimensions for these operators were obtained in \cite{bmsz}
and it was established in \cite{bmsz,bfst}  that the result agreed
with string theory predictions of Arutyunov, Frolov, Russo and Tseytlin
\cite{ft2,ft4,afrt}, in two cases:
The folded string (corresponding to the ground state) and the
circular string (corresponding to an excited state). As we 
explained above, the Inozemtsev chain turns out to be 
intrinsically inconsistent with perturbative BMN scaling, 
starting at four loops.
Just like BMN, the all-loop proposal of Frolov and Tseytlin 
also predicts an expansion in the BMN coupling $\lambda'=\lambda/J^2$. We 
immediately infer that the {\it perturbative} Inozemtsev ansatz therefore has
no chance of reproducing these predictions beyond the three-loop level.
On the other hand, up to three loops BMN scaling is possible, and,
at the two-loop level, proven.
As was shown in the previous chapters, the three-loop
dilatation operator of \cite{bes3} can be emulated by the
Inozemtsev chain. It is therefore clearly very interesting to work out
the Inozemtsev predictions at two and three loops and compare
to string theory. This will be done in the next  sections, for
both the folded and the circular string. At the
two-loop level we can be sure that we are indeed computing the
perturbative gauge theory result. This is in contradistinction to
the three-loop case, where we need to assume the absence of
the terms that break perturbative BMN scaling. Let us recall the string 
result for the two relevant cases of spinning folded and circular
strings. It may be written 
in the following concise way. For the {\it folded} string we have
the parametric expression 
\begin{eqnarray}
\label{eq:string1upper}
\left(\frac{\mathcal{J}_2}{\ellK(t)-\ellE(t)}\right)^2
-\left(\frac{\mathcal{J}_1}{\ellE(t)}\right)^2=\frac{4}{\pi^2}\ ,\\
\label{eq:string1lower}
\left( {\frac{\mathcal{E}}{\ellK(t)}}\right)^2
-\left(\frac{\mathcal{J}_1}{\ellE(t)}\right)^2=\frac{4}{\pi^2}\,t\ ,
\end{eqnarray}
while the {\it circular} string energy ${\cal E}$ is given by
\begin{eqnarray}
\label{eq:string2upper}
\left(\frac{\mathcal{J}_1}{\ellK(t)-\ellE(t)}\right)^2-
\left(\frac{\mathcal{J}_2}{(1-t)\ellK(t)-\ellE(t)}\right)^2
=\frac{4}{\pi^2 t}, \\
\label{eq:string2lower}
\left(\frac{{\cal E}}{\ellK(t)}\right)^2-
\left(\frac{t{\cal J}_1}{\ellK(t)-\ellE(t)}\right)^2
=\frac{4}{\pi^2}(1-t), 
\end{eqnarray}
where ${\cal E},{\cal J}_1,{\cal J}_2$ are
the string energies $E$ and the angular momenta $J_1$,$J_2$
in units of the effective string tension $\lambda$ (alias square root of the
't Hooft coupling), $E=\sqrt{\lambda}~{\cal E}$,
$J_{1,2}=\sqrt{\lambda}~{\cal J}_{1,2}$.
$\ellK(t)$ and $\ellE(t)$ are elliptic integrals of, respectively, the first and second kind.
In both cases the upper equations 
(\ref{eq:string1upper}),(\ref{eq:string2upper}) 
determine the parameter $t$ \footnote{The parameter $t$ is not to be confused with the Inozemtsev coupling constant $t=e^{-2\kappa}$ of the previous chapter.} as a function of
$J_1$,$J_2$, while $E$ is then obtained by substitution of
$t$ into, respectively, the lower equations
(\ref{eq:string1lower}),(\ref{eq:string2lower}). 
These expressions may be expanded for large 
${\cal J}={\cal J}_1+{\cal J}_2$; this results in an 
expansion of the energy as a Taylor series in the BMN coupling
$\lambda'$ through the identification 
({\it cf.}~eq.(\ref{eq:lambda'}))
\begin{equation}
\label{eq:BMN}
\lambda'=\frac{1}{{\cal J}^2}.
\end{equation}
In order to obtain a finite result for the energy
${\cal E}={\cal J} + \ldots$ we should introduce a rescaled
string energy (=lowest charge) ${\cal Q}_2$:
\begin{equation} 
\label{eq:q2}
{\cal Q}_2=\frac{\cal E}{{\cal J}}.
\end{equation}
One then has
\begin{equation}
\label{eq:ftprediction}
{\cal Q}_2=1+\sum_{k=1}^{\infty}~{\cal Q}_2^{(k)}~(\lambda')^k,
\end{equation}
where the expansion coefficients ${\cal Q}_2^{(k)}$
only depend on the
dimensionless ratio (termed ``filling fraction'' below)
\begin{equation}
\label{eq:alpha}
\alpha \equiv \frac{{\cal J}_2}{{\cal J}}=\frac{J_2}{J},
\end{equation}
with $J=J_1+J_2$. Since the expansion eq.(\ref{eq:ftprediction})
is in {\it integer} powers of $\lambda'$, the basic proposal of
Frolov and Tseytlin has been that the $k$-th expansion coefficient
${\cal Q}_2^{(k)}$ should be reproducible by a $k$-loop {\it perturbative}
gauge theory calculation of the anomalous dimension of the operators
eq.(\ref{eq:embryo}). This proposal generalizes the one of BMN, as
the case of ``$J_2$-impurity=$J_2$-magnon'' BMN operators
may be recovered from the folded string solution
eqs.(\ref{eq:string1upper}),(\ref{eq:string1lower}) in the limit
where $J_2$ stays finite ({\it i.e.}~${\cal J}_2 \rightarrow 0$),
but the length of the operator still goes to infinity:
$J \sim J_1 \rightarrow \infty$. In this case
eq.(\ref{eq:ftprediction}) becomes
\begin{equation}
\label{eq:bmn}
{\cal Q}_2=1+\frac{J_2}{J} \left( \sqrt{1+\lambda'}-1 \right)
\end{equation}
{\it i.e.}~the ${\cal Q}_2^{(k)}$ become proportional to the Catalan numbers:
\begin{equation}
\label{eq:bmn2}
{\cal Q}_2^{(k)}=
\frac{J_2}{J}~8~(-\frac{1}{4})^{k+1}~\frac{(2 k-2)!}{k! (k-1)!}.
\end{equation}
As discussed in the introduction, both the BMN and the Frolov-Tseytlin 
proposal make the highly non-trivial assumption that in the AdS/CFT
system two different scaling procedures are nevertheless equivalent:
Recall that in perturbative gauge theory the `t~Hooft coupling
$\lambda$ is {\it small}, while
the string calculation requires $\lambda$ to be {\it large}.
We are now ready to test the Frolov-Tseytlin proposal beyond the
one-loop level, with the help of our long range Bethe ansatz
eq.(\ref{eq:ino-bethe}),(\ref{uphi}).

\subsection{Thermodynamic limit of the Inozemtsev-Bethe equations} 

The thermodynamic limit of the Inozemtsev-Bethe equations is obtained
in close analogy to the procedure employed for the Heisenberg model,
{\it cf.}~\cite{bmsz}. One takes the logarithm of
eq.(\ref{eq:ino-bethe})
and assumes the length $L=J=J_1+J_2$ to be large. This gives
\begin{equation}
\label{eq:discrete}
J~p_j=2 \pi n_j +2 \sum_{\genfrac{}{}{0pt}{}{k=1}{k\neq j}}^{J_2}
\frac{1}{\varphi_j-\varphi_k}.
\end{equation}
In the case of the one-loop Bethe ansatz one works with the 
Bethe roots $u_j$, or, equivalently, the momenta $p_j$. In the
present case one has, in addition to the $\{p_j\}$,
a second set of discrete variables, namely the rapidities 
$\{ \varphi_j \}$. The two sets are related through the
coupling constant dependent
relation eq.(\ref{uphi}). In order to obtain
equations which stay close in form to the original Bethe ansatz
it is most convenient to use eqs.(\ref{eq:discrete}) to
determine the distribution of the rapidities $\{ \varphi_j \}$,
instead of working with the momenta $\{p_j\}$. We can think
of the rapidities as ``deformed'' Bethe roots, since we have
$\varphi_j=u_j + {\cal O}(\lambda)$.

In the limit where the magnon number $M=J_2$ gets large,
the rapidities are expected, after a rescaling, to accumulate along smooth
contours in the complex $\varphi$ plane, just as its undeformed leading order
values $u_j$
\begin{equation}
\label{eq:smooth}
\frac{\varphi_j}{J} \rightarrow \varphi
\qquad \qquad {\rm with} \qquad \qquad
\rho(\varphi)=\frac{1}{J} \sum_{j=1}^{J_2} 
\delta \left(\varphi-\frac{\varphi_j}{J}\right),
\end{equation}
where $\rho(\varphi)$ is a distribution density which
is normalized to the filling fraction
$\alpha=J_2/J$, 
\begin{equation}
\label{eq:norm}
\int_{\mathcal{C}}d\varphi\, \rho(\varphi)=\alpha,
\end{equation}
and $\mathcal{C}$ is the support of the density, {\it i.e.}~the 
union of contours along which the rapidities are distributed. 
The Bethe equations (\ref{eq:discrete}) 
then turn into singular integral equations:
\begin{equation}
\label{eq:BetheCont}
\pint_{\mathcal{C}}\frac{d\varphi'\, 
\rho(\varphi')\, \varphi}{\varphi'-\varphi}
=\frac{1}{2}~\varphi~p(\varphi)+\pi \,
n_{\mathcal{C}(\varphi)} \,\varphi.
\end{equation}
$n_{\mathcal{C}(\varphi)}$ is the mode number at point $\varphi$.
It should be constant along each contour.
Here and in the following the slash through the integral sign
implies a principal part prescription. This equation generalizes
the continuum Bethe equations of \cite{bmsz}. On the right
side the non-trivial potential 
$\varphi~p(\varphi)/2$ appears, which is given by the scaling
limit of the {\it inverse} of the rapidity
$\varphi(p)~p/2$ in eq.(\ref{uphi}). Eq.(\ref{eq:BetheCont})
should be complemented by the momentum conservation condition
$\pint_{\mathcal{C}} d\varphi\,\rho(\varphi)\,p(\varphi)=0$ and
the constraint 
$\pint_{\mathcal{C}} d\varphi\,\rho(\varphi)\,n_{\mathcal{C}(\varphi)}=0$.

We would like to stress that the 
Inozemtsev-Bethe equation eq.(\ref{eq:BetheCont}) is expected to
determine the spectrum of the long-range spin chain (and therefore,
potentially, of the operators in eq.(\ref{eq:embryo})) quite generally
in the regime $J_1,J_2$ large (neglecting, however, 
the ``wrapping'' modes, as explained before). It is {\it not}
necessarily restricted to perturbation theory. The information
on the observables (namely, the commuting charges)
of the long-range spin chain should then be contained in the 
{\it moments} $\bar{Q}_k$ of the rapidity distribution, in generalization
of the result of \cite{as},\cite{emz}:
\begin{equation}
\label{eq:allmoments}
\bar{Q}_k=\frac{1}{2}\int_{\mathcal{C}} 
d\varphi~\frac{\rho(\varphi)}{\varphi^k}.
\end{equation}

However, as far as determining the energy (=lowest charge) of
the spin chain is concerned,
finding the distribution of rapidities is not yet everything.
We also need to uncover the relation between the ``natural'' observables
$\bar{Q}_k$ and the observables we would like to measure,
{\it i.e.}~in particular, the eigenvalue of the dilatation operator.
In the perturbative regime, up to three loops, 
this problem was solved in the last chapter. 
To this order, we are now ready to obtain the anomalous dimension of
the operators eq.(\ref{eq:embryo}) in the two cases of interest
(folded and circular).

\subsection{Folded case}

The folded case corresponds to the ground state of the operators
eq.(\ref{eq:embryo}). The qualitative rapidity distribution is expected
to be as in the one-loop case \cite{bmsz}. We have precisely
two cuts symmetrically placed to both sides of the imaginary axis,
with, respectively, mode numbers $\pm n$. Using the perturbative
result of the last chapter, we find 
(putting the mode number $n$ to one), up to three loops,
the integral equation
\begin{equation}
\label{eq:airfoil}
\pint_a^b d \varphi'~\frac{\rho(\varphi')\,\varphi^2}{\varphi'^2-\varphi^2}
=\frac{1}{4}-\frac{\pi}{2}\,\varphi
+c_1~\frac{\lambda'}{\varphi^2}
+c_2~\frac{\lambda'^2}{\varphi^4},
\end{equation}
which may be treated in a fashion identical to the one
appearing in the multicritical O$(n)$ matrix model \cite{ks}.
The constants $c_1,c_2$ are found from the results of the previous
chapter to be
\begin{equation}
\label{eq:cs}
c_1=\frac{1}{32 \pi^2} \qquad {\rm and} \qquad
c_2=\frac{3}{512 \pi^4}.
\end{equation}
Let us define the moments $\bar{Q}_{2 k}$ of the $\varphi$ distribution by 
\begin{equation}
\label{eq:moments}
\bar{Q}_{2 k}=\int_a^b d\varphi~\frac{\rho(\varphi)}{\varphi^{2 k}} \ .
\end{equation}
The energy $Q_2$ ({\it i.e.}~the anomalous dimensions divided by $J$)
is then determined\footnote{Formally the moments $\bar{Q}_{2 k}$,
and in consequence also the energy $Q_2$, have an 
infinite regular expansion in the BMN coupling $\lambda'$. 
The leading order ${\cal O}(\lambda'^0)$ contribution to
$\bar{Q}_{2 k}$ is nothing but the $2 k$'th one-loop commuting
charge, {\it cf.}~\cite{as}. Obviously
we should discard all terms higher then ${\cal O}(\lambda'^3)$
in $Q_2$ since the thermodynamic
Inozemtsev-Bethe equation eq.(\ref{eq:airfoil})
is accurately describing perturbative gauge theory only to at most 
${\cal O}(\lambda'^3)$. As we discussed, in light of the results
of \cite{dynamic},
we can actually only be completely sure about this validity to 
${\cal O}(\lambda'^2)$ and need to assume the vanishing of two so far
undetermined constants to ensure BMN scaling and, in consequence, validity
of eq.(\ref{eq:airfoil}) to ${\cal O}(\lambda'^3)$.} 
by
\begin{equation}
\label{eq:thermoenergy}
Q_2=1-\frac{\lambda'}{4 \pi^2}~\bar{Q}_2-e_1~\lambda'^2~\bar{Q}_4-
e_2~\lambda'^3~\bar{Q}_6,
\end{equation}
where the constants $e_1,e_2$ were found in the last chapter 
\begin{equation}
\label{eq:es}
e_1=\frac{3}{64 \pi^4} \qquad {\rm and} \qquad
e_2=\frac{5}{512 \pi^6}.
\end{equation}
It is easily verified that these equations are a generalization of
the one-loop ones presented in \cite{bmsz}; their solution proceeds
as before and we will thus be rather brief; for further
details, see \cite{as}. The normalization of the density is
\begin{equation}
\label{eq:norm1}
\int_a^b d\varphi~\rho(\varphi)=-\frac{\alpha}{2}\, .
\end{equation}
with the by now standard trick \cite{bmsz,bfst,as} of analytic continuation
to (formally) negative $J$, resulting in a negative filling fraction
$\alpha=J_2/J<0$ (this flips the complex cuts onto the real axis).
It is again useful to introduce a
generating function for the moments $\bar{Q}_{2 k}$:
\begin{equation}
\label{eq:resolventdef}
H(\varphi)=-\frac{\alpha}{2}+\sum_{k=1}^{\infty}~\bar{Q}_{2 k}~\varphi^{2 k},
\qquad {\rm {\it i.e.}} \qquad
H(\varphi)=
\int_a^b d\varphi'~\rho(\varphi')~\frac{\varphi'^2}{\varphi'^2-\varphi^2} .
\end{equation}
We find the following generalization of the result eq.(2.24) in
\cite{as}:
\begin{eqnarray}
\label{eq:resolvent}
H(\varphi)&=&-\frac{\alpha}{2}+\frac{1}{4}
+c_1~\frac{\lambda'}{\varphi^2} 
+c_2~\frac{\lambda'^2}{\varphi^4}
-\frac{a^2}{b} \sqrt{\frac{b^2-\varphi^2}{a^2-\varphi^2}}
~\Pi\left(-q\frac{\varphi^2}{a^2-\varphi^2},q\right) \\
& & 
-\frac{1}{a b} \sqrt{(b^2-\varphi^2)(a^2-\varphi^2)} \left[
c_1~\lambda'~\frac{1}{\varphi^2}+c_2~\lambda'^2 \left(
\frac{1}{\varphi^4}
+\frac{a^2+b^2}{2 a^2 b^2}\frac{1}{\varphi^2}\right) \right],
\nonumber
\end{eqnarray}
where the modulus $q$ is, as before, $q=1-a^2/b^2$.
The $\Pi$ function is the elliptic integral of the third kind;
for our conventions we refer to \cite{as}.
One then finds that the left edge $a$ of the cut is determined
through the equation
\begin{equation}
\label{eq:edge}
\frac{1}{4}-\ellK(q)~a+c_1~\lambda'~(1-\frac{q}{2})~\frac{1}{a^2}+
c_2~\lambda'^2~(1-q+\frac{3}{8} q^2)~\frac{1}{a^4}
=0,
\end{equation}
while the right edge $b$ is, of course, $b=a/\sqrt{1-q}$.
Finally, one has an equation relating the filling fraction
to the modulus $q$:
\begin{equation}
\label{eq:filling}
\alpha=\frac{1}{2}-\frac{2 a \ellE(q)}{\sqrt{1-q}}+
c_1~\lambda'~\frac{2 \sqrt{1-q}}{a^2}+
c_2~\lambda'^2~\frac{(2-q)\sqrt{1-q}}{a^4}.
\end{equation}
$\ellK(q)$ and $\ellE(q)$ in eqs.(\ref{eq:edge}),(\ref{eq:filling})
are elliptic integrals of the,
respectively, first and second kind.
We are now able to obtain the gauge theory result up 
to three loops, using eq.(\ref{eq:thermoenergy}).This is done
by expanding both the modulus $q=q_0+q_1 \lambda'+q_2 \lambda'^2$
and the interval boundary $a=a_0+a_1 \lambda'+a_2\lambda'^2$ in
powers of $\lambda'$, and eliminating with the help of 
eqs.(\ref{eq:edge}),(\ref{eq:filling}) all constants except for the
leading order modulus $q_0$, which, as before, conveniently parametrizes
the filling fraction $\alpha$ through
\begin{equation}
\label{eq:modulus}
-\alpha= -\frac{J_2}{J}
=\frac{1}{2\sqrt{1-q_0}}\,\frac{\ellE(q_0)}{\ellK(q_0)}-\frac{1}{2}  \ .
\end{equation}
This gives the final, three-loop result
{\small
\begin{eqnarray}
\label{eq:gauge3loop}
Q_2
&=&1+\frac{1}{2 \pi^2} 
\ellK(q_0) (2 \ellE(q_0) - (2 - q_0) \ellK(q_0))~\lambda'
\\
& &+\frac{1}{8 \pi^4}\ellK(q_0)^3 
\left(4 (2-q_0) \ellE(q_0)
-(8-8q_0+q_0^2) \ellK(q_0) \right)~\lambda'^2 \nonumber \\
& &+ \frac{1}{4 \pi^6} \frac
{\ellK(q_0)^5}
{(\ellE(q_0)-\ellK(q_0))(\ellE(q_0)-(1-q_0)\ellK(q_0))} \times \nonumber \\
& & \qquad \times \Big( (8-8 q_0+3q_0^2) \ellE(q_0)^3 \nonumber \\
& & \qquad \qquad 
-(2-q_0)(12-12 q_0+q_0^2)\ellE(q_0)^2\ellK(q_0) \nonumber \\
& & \qquad \qquad \qquad 
+3 (1-q_0)(8-8q_0+q_0^2)\ellE(q_0)\ellK(q_0)^2 \nonumber \\
& & \qquad \qquad \qquad \qquad \qquad
-4(1-q_0)^2(2-q_0)\ellK(q_0)^3 \Big)~\lambda'^3\nonumber
\end{eqnarray}}
We notice that the two-loop (${\cal O}(\lambda'^2)$ contribution 
in eq.(\ref{eq:gauge3loop}) very closely resembles
the eigenvalue of the fourth higher charge $Q_4$, {\it cf.}~eq.(2.25)
in \cite{as}. This finds a natural explanation in terms of
the structure of the energy expression eq.(\ref{eq:thermoenergy}):
Indeed the two-loop energy is ``mostly'' given by the contribution
from the fourth conserved charge of the Heisenberg XXX chain,
with a ``small'' correction, namely the $\lambda'$ correction to the
second Heisenberg charge.

Let us now compare this to the string prediction
eqs.(\ref{eq:string1upper}),(\ref{eq:string1lower}). 
The latter is conveniently rewritten
in Lagrange-inversion form, using eqs.(\ref{eq:BMN}),(\ref{eq:alpha}):
\begin{eqnarray}
\label{eq:lagrange1}
\lambda'&=&\frac{\pi^2}{4 \ellK(t_0)^2}
\left[ \left( \frac{\ellK(t_0)-\ellE(t_0)}{\ellK(t)-\ellE(t)} \right)^2
-\left(\frac{\ellE(t_0)}{\ellE(t)} \right)^2 \right], \\
{\cal Q}_2&=&\frac{\ellK(t)}{\ellK(t_0)} \sqrt{
(1-t)\left(\frac{\ellE(t_0)}{\ellE(t)} \right)^2+
t \left( \frac{\ellK(t_0)-\ellE(t_0)}{\ellK(t)-\ellE(t)} \right)^2 }. 
\nonumber
\end{eqnarray}
where the ``string modulus'' $t_0$ is found to parametrize the
filling fraction through
\begin{equation}
\label{eq:alpha1}
\alpha=1-\frac{\ellE(t_0)}{\ellK(t_0)}.
\end{equation}
One now expands both equations in $t$ around the point $t_0$.
The upper equation gives $\lambda'$ as a power series in 
the variable $(t-t_0)$. Power series inversion yields $t$ as
a power series in $\lambda'$. Finally, substitution of this series
into the expanded lower equation gives ${\cal Q}_2$ as a series in 
$\lambda'$, {\it i.e.}~the sought coefficients in eq.(\ref{eq:ftprediction}). 
We then find, up to third order in $\lambda'$, 
{\small
\begin{eqnarray}
\label{eq:string3loop1}
{\cal Q}_2&=&1 +\frac{2}{\pi^2} \ellK(t_0) 
\left[\ellE(t_0) -(1-t_0) \ellK(t_0) \right]~\lambda' \\
& &+\frac{2}{\pi^4} \ellK(t_0)^3
\left[(1-2 t_0) \ellE(t_0)-(1-t_0)^2 
\ellK(t_0) \right]~\lambda'^2 \nonumber \\
& &+ \frac{4}{\pi^6} \frac
{\ellK(t_0)^5  }
{\ellE(t_0)^2 - 2 (1 - t_0) \ellE(t_0) \ellK(t_0) 
+ (1 - t_0) \ellK(t_0)^2} \times \nonumber \\
& & \qquad \times \Big( (1 - 7 t_0 + 7 t_0^2) \ellE(t_0)^3 
- (1- t_0) (3-14 t_0+7 t_0^2) \ellE(t_0)^2 \ellK(t_0) \nonumber \\
& &\qquad \qquad
+(1-t_0)(3-2 t_0)(1-3 t_0+t_0^2) \ellE(t_0) \ellK(t_0)^2-
(1 - t_0)^4 \ellK(t_0)^3 \Big)~\lambda'^3 \nonumber \\
& &+ {\cal O}(\lambda'^4), \nonumber  
\end{eqnarray}}
In order to compare to gauge theory,
we need to parametrize both results in the same way,
{\it cf.}~eqs.(\ref{eq:modulus}),(\ref{eq:alpha1}). In fact,
we can apply the Gauss-Landen transformation of \cite{bfst}, which
relates the string and gauge theory modulus through
\begin{equation}
\label{eq:GL}
t_0=-\frac{(1-\sqrt{1-q_0})^2}{4\sqrt{1-q_0}}\, .
\end{equation}
to rewrite the three-loop string energy ${\cal Q}_2$ in terms of the
gauge modulus $q_0$:
{\small
\begin{eqnarray}
\label{eq:string3loop2}
{\cal Q}_2
&=&1+\frac{1}{2 \pi^2} 
\ellK(q_0) (2 \ellE(q_0) - (2 - q_0) \ellK(q_0))~\lambda'
\\
& &+\frac{1}{8 \pi^4}\ellK(q_0)^3 
\left(4 (2-q_0) \ellE(q_0)
-(8-8q_0+q_0^2) \ellK(q_0) \right)~\lambda'^2 \nonumber \\
& &+ \frac{1}{16 \pi^6} \frac
{\ellK(q_0)^5}
{(\ellE(q_0)-\ellK(q_0))(\ellE(q_0)-(1-q_0)\ellK(q_0))} \times \nonumber \\
& & \qquad \times \Big( 2(16-16 q_0+7q_0^2) \ellE(q_0)^3 \nonumber \\
& & \qquad \qquad -(2-q_0)(48-48 q_0+7q_0^2)\ellE(q_0)^2\ellK(q_0) \nonumber \\
& & \qquad \qquad \qquad 
+(96-192 q_0+114q_0^2-18q_0^3+q_0^4)\ellE(q_0)\ellK(q_0)^2 \nonumber \\
& & \qquad \qquad \qquad \qquad \qquad
-(1-q_0)(2-q_0)(16-16q_0+q_0^2)\ellK(q_0)^3 \Big)~\lambda'^3\nonumber \\
& &+ {\cal O}(\lambda'^4). \nonumber 
\end{eqnarray}}
Comparing the string theory result eq.(\ref{eq:string3loop2})
to the gauge theory result eq.(\ref{eq:gauge3loop}), we see that
the ${\cal O}(\lambda'^2)$ two-loop terms precisely agree! Our joy 
about this amazing result would
be even greater if this was also the case for the ${\cal O}(\lambda'^3)$
three-loop contribution. However, here we notice, despite a 
close structural resemblance (in particular, the denominators of
the rather involved formulas agree), that the detailed form of the
gauge and string theory results are {\it different}. Furthermore, these
analytic findings explain the numerical results of 
\cite{bfst} for the ground
state gauge energies of the folded case at half filling $\alpha=\frac{1}{2}$.
There the thermodynamic ground state energies of the gauge operators were
estimated by extrapolation of the numerically exact diagonalization
of the dilatation operator for states up to $J=16$. The method showed
excellent gauge-string theory agreement (to about 1\%) at one and two-loops,
but a curious 17\% deviation at three loops. Having at our hands
the exact Inozemtsev-Bethe results eq.(\ref{eq:gauge3loop}) we can
find\footnote{This nice test of the validity of our method was
first suggested to us by N.~Beisert.} the numerically exact values
of the three-loop ground state energies at half-filling\footnote{
It is technically convenient to use the inverse Gauss-Landen transform on
the gauge result eq.(\ref{eq:gauge3loop}); $\alpha=\frac{1}{2}$ 
corresponds to the modulus $t_0=0.8261148$, where
$2 \ellE(t_0)=\ellK(t_0)$. }:
\begin{equation}
\label{eq:gaugenumerical}
Q_2=1+ 0.356016 \lambda'-0.212347 \lambda'^2+0.212147 \lambda'^3,
\end{equation}
while the string result eqs.(\ref{eq:string3loop1}),(\ref{eq:string3loop2})
gives
\begin{equation}
\label{eq:stringnumerical}
{\cal Q}_2=1+ 0.356016 \lambda'-0.212347 \lambda'^2+0.181673 \lambda'^3,
\end{equation}
We see that this {\it explains} the numerical results in 
Table 1 of \cite{bfst}. It shows that the 17\% deviation noticed
there is {\it not} due to numerical inaccuracy of the
extrapolation method -- weak coupling gauge and strong coupling
string theory results really begin to differ at three loops.

The attentive reader might wonder whether the three-loop disagreement 
might disappear when the above constants
eqs.(\ref{eq:cs}),(\ref{eq:es}) are chosen in a different way:
Leaving aside the Inozemtsev chain,
the general form eqs.(\ref{eq:airfoil}),(\ref{eq:thermoenergy})
of ``perturbing'' the Bethe equations of
the one-loop Heisenberg model is of course very natural. 
However, this is not the case. One checks (1) that requiring agreement
between gauge and string theory at two-loops fixes 
$c_1,e_1$ to precisely the values derived from the Inozemtsev-Bethe ansatz,
that furthermore (2) {\it no} choice of $c_2,e_2$ enables one to match the
solution of the deformed Bethe equations and the string prediction,
and that finally (3) the three-loop gauge result 
eq.(\ref{eq:gaugenumerical}) is very sensitive to the precise
values of $c_2,e_2$ and the beautiful agreement with 
table 1 of \cite{bfst},
{\it cf.}~eqs(\ref{eq:gaugenumerical}),(\ref{eq:stringnumerical}) 
is only recovered if we chose the 
``Inozemtsev values'', {\it i.e.}~eqs.(\ref{eq:cs}),(\ref{eq:es}).

On might wonder whether the Inozemtsev chain reproduces
the string result in the non-perturbative BMN regime 
where $\lambda \to \infty$ in conjunction with $J \to \infty$,
while keeping $\lambda'=\lambda/ J^2$ small.
We discussed in section 2.2 that in this case a finite scaled
rapidity exists, {\it cf.}~eq.(\ref{coth}). Upon inversion
we find
\begin{equation}
\label{strong}
\frac{1}{4}~\varphi~p(\varphi)=
\frac{\pi~\varphi}{4 \sqrt{\frac{3}{2} \lambda'}} 
\log 
\frac{1+\sqrt{\frac{3}{2} \lambda'} \frac{1}{2 \pi \varphi}}
{1-\sqrt{\frac{3}{2} \lambda'} \frac{1}{2 \pi \varphi}}.
\end{equation}
In this context the just mentioned finding (2) (no choice of 
$c_2,e_2$ enables one to match the three-loop
string and spin chain results) is unfortunate since
it means that the hyperbolic (i.e.~wrappings are neglected)
Inozemtsev chain at strong coupling also fails to reproduce the
string prediction. Expanding eq.(59) in $\lambda'$ yields
at three loops a potential of the same form 
as found on the right side of eq.(\ref{eq:airfoil}).
One easily checks the quite interesting and non-trivial fact that at 
two loops the strongly coupled hyperbolic chain again reproduces
the same potential as the weakly coupled chain.

For further confirmation of these
findings, let us now turn to the case of the circular string.

\subsection{Circular case}

Here we will be rather brief, as the details are easily filled in
by combining the results and notations of \cite{bmsz,bfst,as} 
and the methodology of the
last section. The Inozemtsev-Bethe equation reads in this case
\begin{equation}
\label{eq:airfoil2}
\pint_c^d d\varphi'\,\tilde{\sigma}(\varphi')\, 
\frac{\varphi^2}{\varphi^2-\varphi'^2}
=\frac{1}{4}- \varphi \log \frac{\varphi+c}{\varphi-c} 
-c_1~\frac{\lambda'}{\varphi^2}+c_2~\frac{\lambda'^2}{\varphi^4},
\end{equation}
where we have rotated the variable $\varphi \rightarrow i \varphi$,
while the three-loop gauge energy $Q_2$ is found from\footnote{
The sign changes in eqs.(\ref{eq:airfoil2}),(\ref{eq:thermoenergy2})
w.r.t.~their folded analogs
eqs.(\ref{eq:airfoil}),(\ref{eq:thermoenergy}) are due to
the rotation $\varphi \rightarrow i \varphi$, and a slightly
different definition of the moments in both cases. See also
\cite{as}.}
\begin{equation}
\label{eq:thermoenergy2}
Q_2=1+\frac{\lambda'}{4 \pi^2}~\bar{Q}_2-e_1~\lambda'^2~\bar{Q}_4+
e_2~\lambda'^3~\bar{Q}_6,
\end{equation}
with the constants $c_1,c_2,e_1,e_2$ again given by
eqs.(\ref{eq:cs}),(\ref{eq:es}). The $\bar{Q}_{2 k}$ are
the moments of the distribution density $\tilde{\sigma}(\varphi)$ and
obtained by
\begin{equation}
\label{eq:moments2}
{\bar Q}_{2 k}=\frac{2}{2 k-1} \frac{1}{c^{2 k-1}} - \int_c^d d\varphi
\frac{\tilde{\sigma}(\varphi)}{\varphi^{2 k}}.
\end{equation}
They are generated by the resolvent
\begin{equation}
\label{eq:resolventdef2}
H(\varphi)=\frac{\alpha}{2}-\sum_{k=1}^{\infty}~{\bar Q}_{2 k}~\varphi^{2k}
\quad {\rm {\it i.e.}} \quad
H(\varphi)=2 c+ \varphi \log \frac{c-\varphi}{c+\varphi} 
+ \int_c^d d\varphi' \tilde{\sigma}(\varphi')
\frac{\varphi'^2}{\varphi'^2-\varphi^2}.
\end{equation}
The normalization is 
\begin{equation}
\label{eq:norm2}
2 c+\int_c^d d\varphi~\tilde{\sigma}(x)=\frac{\alpha}{2},
\end{equation}
where the filling fraction is $\alpha=J_2/J>0$.
The solution involves again the elliptic integral of the third kind,
$\Pi$, with modulus $r=c^2/d^2$ and reads
\begin{eqnarray}
\label{eq:resolvent2}
H(\varphi)&=&\frac{\alpha}{2}-\frac{1}{4}
+c_1~\frac{\lambda'}{\varphi^2} 
-c_2~\frac{\lambda'^2}{\varphi^4}
+\frac{2}{d} \sqrt{(d^2-\varphi^2)(c^2-\varphi^2)}~
\Pi \left(\frac{\varphi^2}{d^2},r\right) \\
& &-\frac{1}{c d}\sqrt{(d^2-\varphi^2)(c^2-\varphi^2)}~\left[
c_1~\lambda'~\frac{1}{\varphi^2}-c_2~\lambda'^2 \left(
\frac{1}{\varphi^4}
+\frac{c^2+d^2}{2 c^2 d^2}\frac{1}{\varphi^2}\right) \right],
\nonumber
\end{eqnarray}
while the conditions fixing the interval boundaries
$c$, $d=c/\sqrt{r}$ and the modulus $r$ are
\begin{equation}
\label{eq:edge2}
-\frac{1}{4}+2\ellK(r)~c+c_1~\lambda'~\frac{1+r}{2}~\frac{1}{c^2}
-c_2~\lambda'^2~\frac{3+2r+3r^2}{8}~\frac{1}{c^4}=0,
\end{equation}
and
\begin{equation}
\label{eq:filling2}
\alpha=\frac{4c}{\sqrt{r}} \left[\left(\sqrt{r}-1 \right) \ellK(r)+
\ellE(r) \right]+
c_1~\lambda'~\frac{(1-\sqrt{r})^2}{c^2}-
c_2~\lambda'^2~\frac{(1-\sqrt{r})^2(3+2\sqrt{r}+3r)}{4 c^4}.
\end{equation}
Expanding the modulus $r=r_0+r_1\lambda'+r_2\lambda'^2$
and the interval boundary $c$, and eliminating all constants except
$r_0$, we find the relation
\begin{equation}
\label{eq:modulus2}
\alpha=\frac{1}{2}-\frac{1}{2 \sqrt{r_0}}+
\frac{1}{2 \sqrt{r_0}} \frac{\ellE(r_0)}{\ellK(r_0)}.
\end{equation}
The three-loop gauge energy is then
{\small
\begin{eqnarray}
\label{eq:circulargauge}
Q_2
&=&1+\frac{2}{\pi^2}
\ellK(r_0) \left( 2\ellE(r_0)-(1-r_0)\ellK(r_0)\right)~\lambda' \\
& &-\frac{2}{\pi^4} 
\ellK(r_0)^3 \left(
4(1+r_0)\ellE(r_0)-(1-r_0)(3+r_0)\ellK(r_0) \right)~\lambda'^2
\nonumber \\
& &+\frac{16}{\pi^6} 
\frac{\ellK(r_0)^5}{\ellE(r_0)\left( \ellE(r_0)-(1-r_0)\ellK(r_0)
\right)} \times \nonumber \\
& & \qquad \times \Big( (3+2r_0+3r_0^2)\ellE(r_0)^3 
-(1-r_0)(8+3r_0+r_0^2)\ellE(r_0)^2 \ellK(r_0) \nonumber \\
& & \qquad \qquad \qquad \qquad 
+(1-r_0)^2(7+r_0) \ellE(r_0) \ellK(r_0)^2
-2 (1-r_0)^3 \ellK(r_0)^3 \Big)~\lambda'^3 \nonumber
\end{eqnarray}}
Now we will again compare this to the string prediction, 
eqs.(\ref{eq:string2upper}),(\ref{eq:string2lower}). 
Written in Lagrange-inversion form, we have
\begin{eqnarray}
\label{eq:lagrange2}
\lambda'&=&\frac{\pi^2 t}{4 t_0^2 \ellK(t_0)^2}
\left[ \left( \frac{\ellK(t_0)-\ellE(t_0)}{\ellK(t)-\ellE(t)} \right)^2
-\left(\frac{\ellE(t_0)-(1-t_0)\ellK(t_0)}{\ellE(t)-(1-t)\ellK(t)}
\right)^2 \right], \\
{\cal Q}_2&=&\frac{t \ellK(t)}{t_0 \ellK(t_0)} \sqrt{
\frac{1}{t}\left( \frac{\ellK(t_0)-\ellE(t_0)}{\ellK(t)-\ellE(t)} \right)^2-
\frac{1-t}{t} 
\left(\frac{\ellE(t_0)-(1-t_0)\ellK(t_0)}{\ellE(t)-(1-t)\ellK(t)}
\right)^2 }, 
\nonumber
\end{eqnarray}
where now the string modulus $t_0$ is parametrizing the
filling fraction through
\begin{equation}
\label{eq:alpha2}
\alpha=1-\frac{1}{t_0} + \frac{1}{t_0}~\frac{\ellE(t_0)}{\ellK(t_0)}.
\end{equation}
As before, power series inversion and substitution gives, from 
eqs.(\ref{eq:lagrange2}),
{\small
\begin{eqnarray}
\label{eq:circularstring}
{\cal Q}_2
&=&1+\frac{2}{\pi^2} \ellE(t_0) \ellK(t_0)~\lambda' \\
& &-\frac{2}{\pi^4}
\ellK(t_0)^3 \left( (2-t_0)\ellE(t_0)-(1-t_0) \ellK(t_0)
\right)~\lambda'^2 \nonumber \\
& &+\frac{4}{\pi^6} \frac{\ellK(t_0)^5}
{\ellE(t_0)^2-(1-t_0)\ellK(t_0)^2} \times \nonumber \\
& & \qquad \times \Big( (7-7t_0+t_0^2)\ellE(t_0)^3
-7(2-t_0)(1-t_0) \ellE(t_0)^2 \ellK(t_0) \nonumber \\
& & \qquad \qquad 
+(1-t_0)(9-9t_0+t_0^2)\ellE(t_0)\ellK(t_0)^2
-(2-t_0)(1-t_0)^2 \ellK(t_0)^3 \Big)~\lambda'^3 \nonumber \\
& & + {\cal O}(\lambda'^4) \nonumber
\end{eqnarray}}
After the Gauss-Landen transformation
\begin{equation}
\label{eq:GLc}
t_0=-\frac{4\sqrt{r_0}}{(1-\sqrt{r_0})^2}
\end{equation}
the string energy up to three-loop order becomes
{\small
\begin{eqnarray}
\label{eq:circularstring2}
{\cal Q}_2
&=&1+\frac{2}{\pi^2}
\ellK(r_0) \left( 2\ellE(r_0)-(1-r_0)\ellK(r_0)\right)~\lambda' \\
& &-\frac{2}{\pi^4} 
\ellK(r_0)^3 \left(
4(1+r_0)\ellE(r_0)-(1-r_0)(3+r_0)\ellK(r_0) \right)~\lambda'^2
\nonumber \\
& &+\frac{4}{\pi^6} 
\frac{\ellK(r_0)^5}{\ellE(r_0)\left( \ellE(r_0)-(1-r_0)\ellK(r_0)
\right)} \times \nonumber \\
& & \qquad \times \Big( 2 (7+2r_0+7r_0^2)\ellE(r_0)^3 
-(1-r_0)(35+6r_0+7r_0^2)\ellE(r_0)^2 \ellK(r_0) \nonumber \\
& & \qquad \qquad \qquad \qquad 
+(1-r_0)^2(29+2 r_0+r_0^2) \ellE(r_0) \ellK(r_0)^2
-8 (1-r_0)^3 \ellK(r_0)^3 \Big)~\lambda'^3 \nonumber \\
& & + {\cal O}(\lambda'^4) \nonumber
\end{eqnarray}}
The result of the comparison between the string theory prediction
eq.(\ref{eq:circularstring2})
and the gauge theory computation eq.(\ref{eq:circulargauge})
leads to the same conclusion as the previously treated case of
the folded string. There is a beautiful and non-trivial
agreement at the two-loop order; on the other hand, the three-loop terms,
despite great structural similarity, are definitely different.
Again, varying the constants $c_2,e_2$ 
in eqs.(\ref{eq:airfoil2}),(\ref{eq:thermoenergy2}) does not help. 

It is instructive to consider the special case
of half-filling $\alpha=1/2$. Here the solution 
becomes algebraic, and the string result reads, to all orders
\begin{equation}
\label{eq:halfstring}
{\cal
  Q}_2=\sqrt{1+\lambda'}=1+\frac{\lambda'}{2}-\frac{\lambda'^2}{8}+
\frac{\lambda'^3}{16}+\ldots
\end{equation}
The three-loop gauge theory result, on the other hand, under
the assumption of the validity of three-loop perturbative BMN scaling,
yields
\begin{equation}
\label{eq:halfgauge}
Q_2=1+\frac{\lambda'}{2}-\frac{\lambda'^2}{8}+0~\lambda'^3 
\end{equation}
The three-loop contribution to the ``circular'' state {\it vanishes}!

\subsection{A curious observation}

It is clearly of some interest to pin down the
difference between three-loop perturbation theory, and the
string prediction. For the folded case we have
\begin{equation}
\label{eq:diff1}
Q_2-{\cal Q}_2=-\frac{q_0^2}{16 \pi^6} \ellK(q_0)^5
\left( 2 \ellE(q_0)-(2-q_0) \ellK(q_0) \right)~\lambda'^3,
\end{equation}
and in the circular case
\begin{equation}
\label{eq:diff2}
Q_2-{\cal Q}_2=-\frac{4(1-r_0)^2}{\pi^6} \ellK(r_0)^5
\left( 2 \ellE(r_0)-(1-r_0) \ellK(q_0) \right)~\lambda'^3.
\end{equation}
Curiously, this looks like a non-linear (and therefore, from
the point of view of the perturbative spin chain, non-local)
contribution to the energy. In {\it both} cases the difference is 
proportional to the product of the one-loop second moment 
$\bar{Q}_2^{(1)}$
and its leading correction $\bar{Q}_2^{(2)}$:
\begin{equation}
\label{eq:mystery}
Q_2-{\cal Q}_2=-\frac{\lambda'^3}{32 \pi^4}~\bar{Q}_2^{(1)}~\bar{Q}_2^{(2)}
\qquad {\rm where} \qquad
{\bar Q}_2=\bar{Q}_2^{(1)}+\lambda' \bar{Q}_2^{(2)}+{\cal
  O}(\lambda'^2)
\end{equation}
But the significance or wider validity of this observation
remains unclear to us at the moment.
It would be interesting to see whether the same ``recipe''
also allows to account for the three-loop near-plane wave discrepancy
of Callan {\it et al.}~\cite{callan}.

\bigskip
\leftline{\bf Acknowledgments}

\noindent We would like to thank 
Denis Bernard, Mike Douglas, C\'esar G\'omez, Volodya Kazakov, Ivan Kostov, 
Stefano Kovacs, Charlotte Kristjansen, Esperanza Lopez, Jan Plefka, 
Jorge Russo, Emery Sokatchev, Bogdan Stefanski, Stefan Vandoren and, 
especially,
Gleb Arutyunov and Niklas Beisert for very useful discussions.
D.S.~and M.S. thank, respectively, AEI Potsdam and
CEA Saclay for hospitality. This research is supported in
part by the European network EUCLID, HPRN-CT-2002-00325.

\end{document}